\documentclass[10pt,compsocconf]{IEEEtran}
\usepackage{algorithm,amsbsy,amsmath,amssymb,epsfig,bbm,mathrsfs,multirow,amsthm,xcolor}
\usepackage{epstopdf}
\usepackage{graphicx,caption,subcaption}
\usepackage{setspace}
\usepackage{algorithmic}
\usepackage{subcaption,float}
\usepackage[hyphens]{url}
\usepackage[hidelinks]{hyperref}
\hypersetup{breaklinks=true}
\begin{document}
\title{Energy Demand Prediction with Federated Learning for Electric Vehicle Networks}
\author{{\IEEEauthorblockN{Yuris Mulya Saputra$^1$, Dinh Thai Hoang$^1$, Diep N. Nguyen$^1$, Eryk Dutkiewicz$^1$,\\
Markus Dominik Mueck$^2$, and Srikathyayani Srikanteswara$^2$\\}
$^1$ School of Electrical and Data Engineering, University of Technology Sydney (UTS), Australia\\
$^2$ Intel Corporation\\ \vspace{-5mm}}
\thanks{This work has been supported in part by Intel's University Research Office.}}

\maketitle
\thispagestyle{empty}
\begin{abstract}
In this paper, we propose novel approaches using state-of-the-art machine learning techniques, aiming at predicting energy demand for electric vehicle (EV) networks. These methods can learn and find the correlation of complex hidden features to improve the prediction accuracy. First, we propose an energy demand learning (EDL)-based prediction solution in which a charging station provider (CSP) gathers information from all charging stations (CSs) and then performs the EDL algorithm to predict the energy demand for the considered area. However, this approach requires frequent data sharing between the CSs and the CSP, thereby driving communication overhead and privacy issues for the EVs and CSs. To address this problem, we propose a federated energy demand learning (FEDL) approach which allows the CSs sharing their information without revealing real datasets. Specifically, the CSs only need to send their trained models to the CSP for processing. In this case, we can significantly reduce the communication overhead and effectively protect data privacy for the EV users. To further improve the effectiveness of the FEDL, we then introduce a novel clustering-based EDL approach for EV networks by grouping the CSs into clusters before applying the EDL algorithms. Through experimental results, we show that our proposed approaches can improve the accuracy of energy demand prediction up to $24.63$\%  and decrease communication overhead by $83.4\%$ compared with other baseline machine learning algorithms.	

\end{abstract}

{\it Keywords-} Energy demand, electric vehicle, charging station, federated learning, clustering.

\maketitle

\IEEEdisplaynontitleabstractindextext

\IEEEpeerreviewmaketitle

\section{Introduction}\label{sec:Int}

The electric vehicles (EVs) have emerged as one of the sustainable solutions to transform the conventional transportation systems with less oil use, high energy efficiency, and low gas emissions. According to International Energy Agency~\cite{Cnbc:2019}, the number of EVs on the road will rise tremendously by more than 4000\% in 2030. This trend will lead to an explosion of energy demand, i.e., energy consumption for EVs' batteries, in the power market and make a significant impact on the power grid system. 

Generally, the power grid supplies the energy for charging stations (CSs) once receiving requests from EVs~\cite{Tang:2014}. However, this approach experiences a serious energy transfer congestion when a huge number of EVs charging the energy simultaneously~\cite{Lopes:2011} and lead to high energy transfer cost for the charging station provider (CSP). To deal with this problem, energy can be reserved in advance at the CSs to meet real-time demands from the EVs~\cite{You:2014}. Particularly, in~\cite{You:2014}, the authors proposed a solution to orchestrate EVs' charging using offline scheduling scheme. However, without considering the real demand history, this approach may suffer from under/over-utilization, due to the unpredictable and dynamic EVs' energy demands. Hence, effective and intelligent approaches to predict energy demands for CSs in the EV networks are of significance to optimize energy efficiency. As such, the CSP can reduce the energy transfer cost and stabilize the energy demand in the power grid system. Furthermore, the CSP can provide real-time energy charging and guarantee stable energy supply for the upcoming EVs. 

To predict energy demand for EV networks, the authors in~\cite{Majidpour:2015} introduced a mobile application which can estimate energy availability at each CS using k-nearest neighbor (kNN) algorithm. Specifically, the history of consumed energy and charging duration for connected EVs is captured as the dataset and then stored on the remote server. In~\cite{Chis:2017}, the authors proposed a reinforcement learning-based demand response scheme to optimize the amount of energy charging for an individual EV based on daily forecasted price policy. Furthermore, the authors in~\cite{Fukushima:2018} developed a driving activity-based recommendation system applying a multiple regression-based learning approach, aiming at improving the energy consumption prediction accuracy for EVs. In~\cite{Ma:2017}, an online learning to optimize EVs' charging demands using previous-day pricing profiles from the distribution company was proposed. Alternatively, the authors in~\cite{Lopez:2019} designed a smart charging policy for EVs using machine learning tools including deep neural network (DNN), shallow neural network (SNN), and kNN to decide the charging time when the EVs are connected to the CSs. Nonetheless, these approaches only consider energy demand prediction independently at each EV or CS, and thus they may not be effective for the whole EV network. In fact, predicting the energy demand using shared information or global models can further improve the prediction accuracy. Furthermore, those machine learning-based solutions do not address the communication overhead and privacy issues, which are very crucial in the future EV networks.


In this paper, we introduce state-of-the-art machine learning-based approaches which can not only significantly improve the accuracy of energy demand prediction, but also remarkably reduce the communication overhead for EV networks. In particular, we first introduce a communication model using the CSP as a centralized node to gather all information from the CSs in a considered area. We then develop an energy demand  learning (EDL)-based solution utilizing deep learning method to help the CSP accurately predict energy demands for the CSs in this area. However, this approach requires the CSs to share their local data with the CSP, and thus it may suffer from serious overhead and privacy issues. To address these issues, we propose a novel federated energy demand learning (FEDL) approach in which the CSs only need to share their trained models obtained from their datasets instead of sharing their real datasets. To further improve the prediction accuracy, we develop the clustering-based EDL approach which can classify the CSs into several clusters before the learning process is performed. In this way, we can reduce the dimensionality of the dataset based on the useful feature classification~\cite{He:2019}, and thus the biased prediction can be minimized~\cite{Li:2018}. Through experimental results, we show that our proposed approaches can improve the accuracy of energy demand prediction up to $24.63$\% and significantly reduce the communication overhead by $83.4\%$ compared with those of other baseline machine learning algorithms. The major contributions are summarized as follows:

\begin{itemize}
\item We design the state-of-the-art machine learning-based approach that leverages the EDL algorithm to improve the accuracy of energy demand prediction for CSs through the CSP.

\item We introduce the novel approach using FEDL method to reduce overhead and privacy for the CSs.

\item We develop the clustering-based EDL approach to minimize the cost of biased prediction, thereby further improving the prediction accuracy.
	
\item We conduct extensive experimental results to evaluate the efficiency of the proposed methods using the real CS session dataset in Dundee city, the United Kingdom.

\end{itemize}



\begin{figure}[!t]
	\centering
	\includegraphics[scale=0.31]{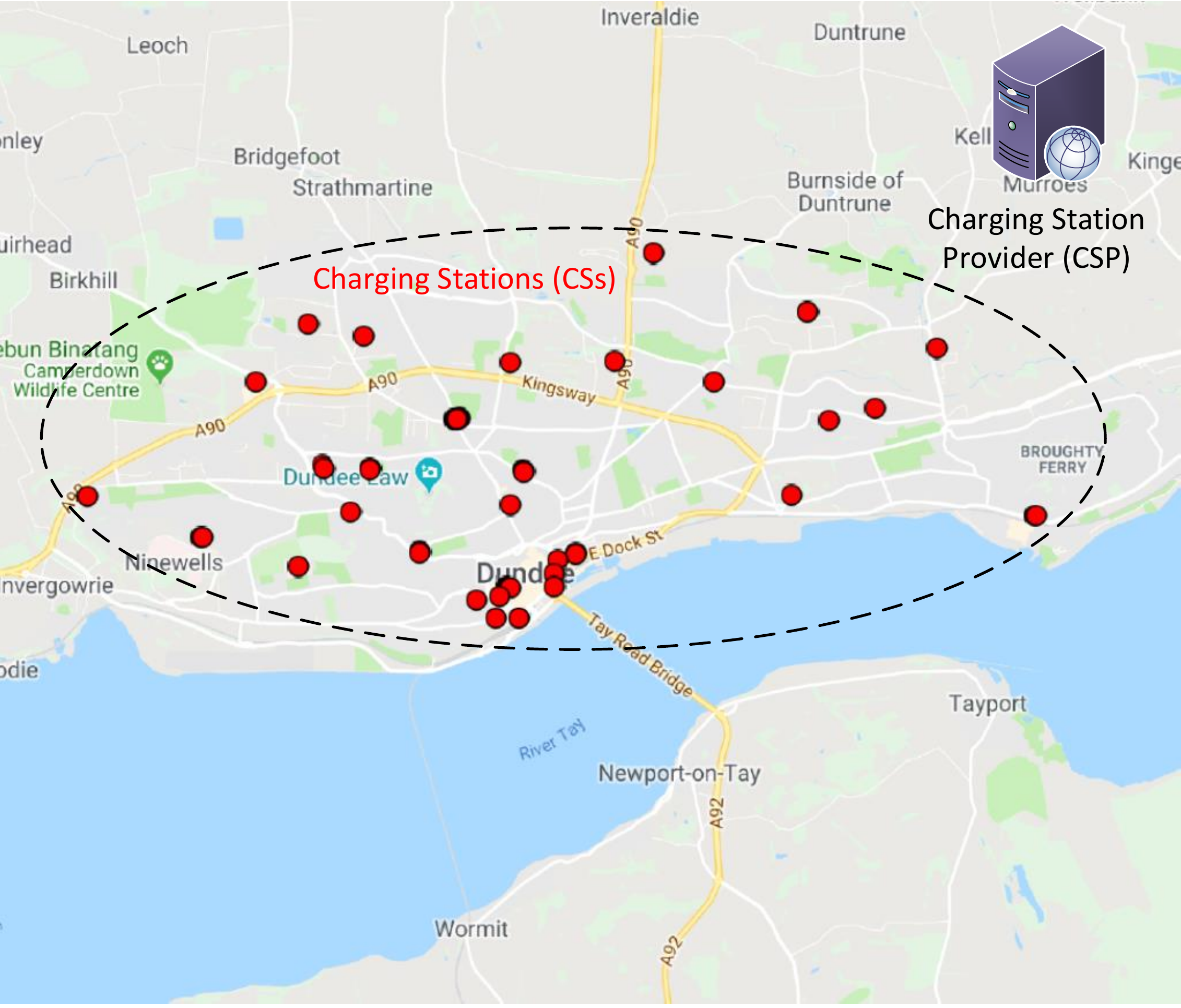}
	\caption{Charging station distribution in Dundee city, the United Kingdom between 2017 and 2018~\cite{Dundee:2018}.}
	\label{fig:Edge_Architecture}
\end{figure}

\section{System Model}
\label{sec:SM}
	
The distribution of CSs is described in Fig.~\ref{fig:Edge_Architecture}. This distribution is obtained from real data provided in~\cite{Dundee:2018} for locations of the CSs in Dundee City, the United Kingdom between 2017 and 2018. Each CS receives energy from power sources through power grid system to serve EVs. Each CS has a log file to record all transactions of EVs charging at this station. In particular, each transaction includes the following information: CS ID, EV ID, charging date, charging time, and consumed energy within a particular period. Then, the CS can store each transaction in its log file and use the log file to predict energy demand in the next period. This information will be captured and updated in the log file periodically. Since the number of transactions at each CS is usually very few, and thus not sufficient to predict the energy demand accurately. As a result, the CSP in the EV network will take responsibilities to gather information, i.e., transactions or trained models, from the CSs and then perform learning algorithms to predict energy demand for the whole network. In this way, the CS can reduce the energy cost to serve the upcoming EVs as the energy stored in advance may have inexpensive price~\cite{Roman:2011}. Additionally, we can provide real-time energy charging and guarantee stable energy supply for the EVs in the EV network.

Let $\mathcal{I} = \{1,\ldots,i,\ldots,I\}$ denote the set of CSs in the EV network. 
We also define $\mathcal{M}_i = \{1,\ldots,m,\ldots,M_i\}$, as the set of charging transactions at CS-$i$. We denote energy demand for transaction $m$ at CS-$i$ as $\omega_{i}^{m}$.

\section{Energy Demand Prediction} \label{sec:PFT}
In this section, we present three approaches to predict energy demands of CSs including centralized EDL, FEDL, and clustering-based EDL algorithms. Specifically, each approach is useful to implement in the following particular scenarios.


\subsection{Centralized Energy Demand Learning}
\label{subsec:PF1}


This method is especially applicable when the CSs have limited computing resources and cannot execute the EDL process by themselves. Specifically, in this approach, the CSP first collects all information, i.e., log files $\mathbf{X}_i, \forall i \in \mathcal{I}$, from the CSs, in order to make an accumulated log file $\mathbf{X}_{\emph{\mbox{csp}}}$. Then, we develop a deep learning algorithm to help the CSP predict the energy demand in the EV network with high accuracy. To learn through the DNN in the CSP, we denote $\mathcal{L} = \{1,\ldots,l,\ldots,L\}$ to be the set of learning layers. In particular, the $\mathbf{X}_{\emph{\mbox{csp}}}$ of layer $l$, i.e., $\mathbf{X}_{\emph{\mbox{csp}}}^l$, is used as the training input matrix to produce the following output matrix of layer-$l$ in the CSP
\begin{equation}
\label{eqn3a}
\begin{aligned}
\mathbf{Y}_{\emph{\mbox{csp}}}^l = a_{\emph{\mbox{csp}}} \big(\mathbf{G}_l\mathbf{X}_{\emph{\mbox{csp}}}^l + \mathbf{h}_l\big),
\end{aligned}
\end{equation}
where $\mathbf{G}_l$ and $\mathbf{h}_l$ are the global weight matrix and global bias vector of the layer $l$, respectively, while $a_{\emph{\mbox{csp}}}$ is a \emph{tanh} activation function~\cite{Zhang2:2018} to compute the hyperbolic tangent of $\mathbf{X}_{\emph{\mbox{csp}}}^l$, which is expressed by
\begin{equation}
\label{eqn3a2}
\begin{aligned}
a_{\emph{\mbox{csp}}} = \frac{e^{\mathbf{X}_{\emph{\mbox{csp}}}^l}-e^{-\mathbf{X}_{\emph{\mbox{csp}}}^l}}{e^{\mathbf{X}_{\emph{\mbox{csp}}}^l}+e^{-\mathbf{X}_{\emph{\mbox{csp}}}^l}}.
\end{aligned}
\end{equation}
When hidden layers are applied, we can define $\mathbf{X}_{\emph{\mbox{csp}}}^{l+1} = \mathbf{Y}_{\emph{\mbox{csp}}}^l$. Furthermore, we apply a dropout layer $l_{\emph{\mbox{drop}}}$ after the last hidden layer to avoid the generalization error and overfitting issue by randomly dropping the $\mathbf{X}_{\emph{\mbox{csp}}}^{l_{\emph{\mbox{drop}}}}$ with a fraction rate $f$. Hence, the remaining input elements are scaled by $\frac{1}{1-f}$. 

Suppose that $\boldsymbol{\upsilon} = (\mathbf{G}, \mathbf{h})$, where $\mathbf{G} = [\mathbf{G}_1,\ldots,\mathbf{G}_l,\ldots,\mathbf{G}_L]$ and $\mathbf{h} = [\mathbf{h}_1,\ldots,\mathbf{h}_l,\ldots,\mathbf{h}_L]$, as the global model for all layers, the prediction error $\rho(\boldsymbol{\upsilon}^{(\phi)})$ for epoch time $\phi$ (i.e., the time when all transactions of $\mathbf{X}_{\emph{\mbox{csp}}}$ has been observed) in the CSP is expressed as follows:
\begin{equation}
\label{eqn3b}
\begin{aligned}
\rho(\boldsymbol{\upsilon}^{(\phi)}) = \sum_{i=1}^{I}\sum_{m=1}^{M_i}\rho^{m}_{i}(\boldsymbol{\upsilon}^{(\phi)}), 
\end{aligned}
\end{equation}
where $\rho^{m}_{i}(\boldsymbol{\upsilon}^{(\phi)}) = (y_{\emph{\mbox{i}}}^{m} - x_{\emph{\mbox{i}}}^{m})^2$. In this case, $x_{\emph{\mbox{i}}}^{m} = \omega_{i}^{m}$ and $y_{\emph{\mbox{i}}}^{m}$ are the elements of input matrix $\mathbf{X}_{\emph{\mbox{csp}}}^1$ and output matrix $\mathbf{Y}_{\emph{\mbox{csp}}}^L$, respectively. Then, we can compute the global gradient of using EDL at $\phi$ by 
\begin{equation}
\label{eqn3c}
\begin{aligned}
\nabla \boldsymbol{\upsilon}^{(\phi)} = \frac{\partial \rho(\boldsymbol{\upsilon}^{(\phi)})}{\partial \boldsymbol{\upsilon}^{(\phi)}}.
\end{aligned}
\end{equation}

After $\nabla \boldsymbol{\upsilon}^{(\phi)}$ is obtained, the CSP updates the global model $\boldsymbol{\upsilon}^{(\phi)}$ to minimize the prediction error, i.e., $\underset{\boldsymbol{\upsilon}}{\text{\bf min }}\rho(\boldsymbol{\upsilon})$, using adaptive learning rate optimizer \emph{Adam}~\cite{Kingma:2015} which produces fast convergence and significant robustness to the model. Let $\eta_\phi$ and $\delta_\phi$ denote the exponential moving average of the $\nabla \boldsymbol{\upsilon}^{(\phi)}$ and the squared $\nabla \boldsymbol{\upsilon}^{(\phi)}$ to infer the variance at $\phi$, respectively. Then, the update rules of $\eta_{\phi+1}$ and $\delta_{\phi+1}$ can be described as follows:
\begin{equation}
\begin{aligned}
\label{eqn3d1}
\eta_{\phi+1} &= \gamma_\eta^\phi \eta_{\phi} + (1 - \gamma_\eta^\phi)\nabla \boldsymbol{\upsilon}^{(\phi)},\\
\delta_{\phi+1} &= \gamma_\delta^\phi \delta_{\phi} + (1 - \gamma_\delta^\phi)(\nabla \boldsymbol{\upsilon}^{(\phi)})^2,
\end{aligned}
\end{equation}
where $\gamma_\eta^\phi$ and $\gamma_\delta^\phi \in [0,1)$ indicate the exponential decay steps of $\eta_\phi$ and  $\delta_\phi$ at $\phi$, respectively. 
Furthermore, we consider the learning step $\lambda$ to determine how fast we need to update the global model. Specifically, the update rule for $\lambda$ can be expressed by
\begin{equation}
\label{eqn3d}
\begin{aligned}
\lambda_{\phi+1} = \lambda\frac{\sqrt{1 - \gamma_\delta^{\phi+1}}}{1 - \gamma_\eta^{\phi+1}}.
\end{aligned}
\end{equation}
Then, the global model $\boldsymbol{\upsilon}_k^{(\phi+1)}$ to learn $\mathbf{X}_{\emph{\mbox{csp}}}$ for the next epoch time $\phi+1$ is updated as follows:
\begin{equation}
\label{eqn3e}
\begin{aligned}
\boldsymbol{\upsilon}^{(\phi+1)} = \boldsymbol{\upsilon}^{(\phi)} - \lambda_{\phi+1}\frac{\eta_{\phi+1}}{\sqrt{\delta_{\phi+1}} + \varepsilon},
\end{aligned}
\end{equation}
where $\varepsilon$ represents a constant to avoid zero division when the $\sqrt{\delta_{\phi+1}}$ is almost zero. The learning process repeats and then terminates when the prediction error converges, or a certain number of epoch time $T$ is reached. In this case, the final global model $\boldsymbol{\upsilon}^*$ in the CSP is obtained to predict $\mathbf{\hat Y}_{\emph{\mbox{csp}}}$ of training dataset $\mathbf{X}_{\emph{\mbox{csp}}}$ and new dataset $\mathbf{\hat X}_{\emph{\mbox{csp}}}$ using Eq.~(\ref{eqn3a}).  The algorithm for energy demand prediction using EDL is summarized in Algorithm~\ref{CDL-PC}. The processes between Lines 4 and 11 are implemented in the CSP.

\begin{algorithm}[]
	\caption{EDL-Based Prediction Algorithm} \label{CDL-PC}
	
	\begin{algorithmic}[1] 
		\STATE Initialize $a_{\emph{\mbox{csp}}}$ and $\boldsymbol{\upsilon}^{(\phi)}$ when $\phi=0$
		
		\STATE Generate $\mathbf{X}_i, \forall i \in \mathcal{I}$
		
		\STATE Send $\mathbf{X}_i, \forall i \in \mathcal{I}$ to the CSP
		
		\STATE Set $\mathbf{X}_{\emph{\mbox{csp}}}^1 \leftarrow \underset{i \in \mathcal{I}}{\sum}\mathbf{X}_i$ with $e^{m}_i$,  $\forall m \in \mathcal{M}_i, \forall i \in \mathcal{I}$
		
		\WHILE{$\phi \leq T \text{ {\bf and} } \rho(\boldsymbol{\upsilon}^{(\phi)}) \text{ do not converge}$}
		
		\STATE Learn $\mathbf{X}_{\emph{\mbox{csp}}}^1$ to obtain $\mathbf{Y}_{\emph{\mbox{csp}}}^L$ at layer-$L$ using $\boldsymbol{\upsilon}^{(\phi)}$
		
		\STATE Calculate $\rho(\boldsymbol{\upsilon}^{(\phi)})$ and $\nabla \boldsymbol{\upsilon}^{(\phi)}$
		
		\STATE $\phi \leftarrow \phi+1$
		
		\STATE Update $\boldsymbol{\upsilon}^{(\phi)}$
		
		\ENDWHILE
		
		\STATE Predict $\mathbf{\hat Y}_{\emph{\mbox{csp}}}$ using $\mathbf{\hat X}_{\emph{\mbox{csp}}}$ and $\boldsymbol{\upsilon}^*$ in the CSP for the next-period energy requests
		
	\end{algorithmic}
\end{algorithm}

\subsection{Federated Energy Demand Learning}
\label{subsec:PF2}

\begin{figure}[!t]
	\centering
	\includegraphics[scale=0.32]{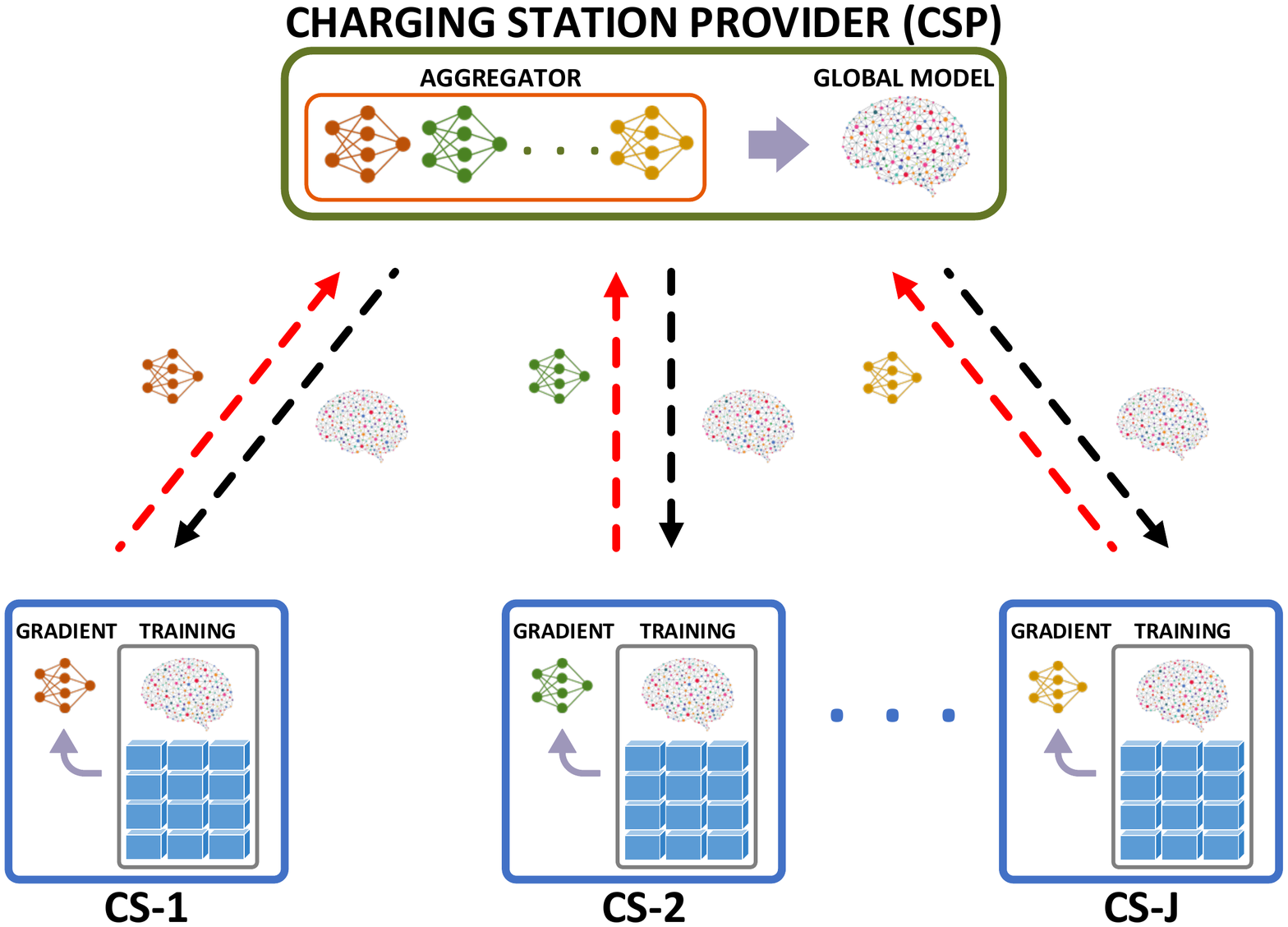}
	\caption{Federated energy demand learning architecture.}
	\label{fig:DDL_Architecture}
\end{figure}

Since the CSP needs to collect data from CSs, this centralized learning may lead to the communication overhead and data privacy concerns. To deal with these issues, we develop a framework using federated energy demand learning (FEDL) method. Particularly, the CSP only requires to collect the trained models, i.e., gradient information, from the set of CSs, and then updates the global model efficiently before sending back to the CSs~\cite{Dean:2012}. After that, the CSs can use this information to learn by themselves using deep learning method. Given that $\mathcal{J} = \{1,\ldots,j,\ldots,J\}$ as the set of CSs which acts as workers to implement the EDL algorithms using their $\mathbf{X}_j$ locally as illustrated in Fig.~\ref{fig:DDL_Architecture}. Thus, $\mathbf{X}_j$ is expressed by
\begin{equation}
\label{eqn3f0}
\begin{aligned}
\mathbf{X}_j = \sum_{i=1}^{I}\beta_j^i\mathbf{X}_i,
\end{aligned}
\end{equation}
where $\beta_j^i$ is a unique binary variable with $\beta_j^i=1$ indicating that the CS-$j$ has information from CS-$i$, and $\beta_j^i=0$ otherwise.

For DNN, the output matrix of layer-$l$ at CS-$j$ is generated by 
\begin{equation}
\label{eqn3f}
\begin{aligned}
\mathbf{Y}_{\emph{\mbox{j}}}^l = a_{\emph{\mbox{j}}} \big(\mathbf{G}_l\mathbf{X}_j^l + \mathbf{h}_l\big),
\end{aligned}
\end{equation}
where $\mathbf{X}_j^l$ is the input matrix of layer $l$ at CS-$j$ (with $\mathbf{X}_j^1 = \mathbf{X}_j$ and $a_{\emph{\mbox{j}}}$ is the \emph{tanh} activation function at CS-$j$. Additionally, we use dropout layer to eliminate the input $\mathbf{X}_j^{l_{\emph{\mbox{drop}}}}$ by a fraction rate $f$. At the output layer, we can obtain output matrix $\mathbf{Y}_{\emph{\mbox{j}}}^L$ and find the prediction error for each $\phi$ at CS-$j$ as follows:
\begin{equation}
\label{eqn3g}
\begin{aligned}
\rho_j(\boldsymbol{\upsilon}^{(\phi)}) = \sum_{i=1}^{I}\sum_{m=1}^{M_i}\beta_j^i\rho^{m}_{i}(\boldsymbol{\upsilon}^{(\phi)}).
\end{aligned}
\end{equation}

Next, we can compute the local gradient at CS-$j$ by 
\begin{equation}
\label{eqn3h}
\begin{aligned}
\nabla \boldsymbol{\upsilon}_j^{(\phi)} = \frac{\partial \rho_j(\boldsymbol{\upsilon}^{(\phi)})}{\partial \boldsymbol{\upsilon}^{(\phi)}}.
\end{aligned}
\end{equation}
Upon completing $\nabla \boldsymbol{\upsilon}_j^{(\phi)}$ for every $\phi$, each CS-$j$ sends the local gradient to the CSP for global gradient aggregation as expressed by
\begin{equation}
\label{eqn3i}
\begin{aligned}
\nabla \boldsymbol{\upsilon}^{(\phi)} = \frac{1}{J}\sum_{j=1}^J \nabla \boldsymbol{\upsilon}_i^{(\phi)}.
\end{aligned}
\end{equation}
In particular, the CSP acts as a model server to accumulate the gradients, and then updates the global model $\boldsymbol{\upsilon}^{(\phi)}$ before sending back to the CS-$j$, $\forall j \in \mathcal{J}$ (as illustrated in Fig.~\ref{fig:DDL_Architecture}). This allows all CS-$j$, $\forall j \in \mathcal{J}$ to collaborate by sharing local model information to each other to further improve the prediction accuracy through the CSP. To ensure that the gradient staleness is 0, the gradient aggregation is enabled immediately after $J$ local gradients are received by the CSP synchronously. In this way, the gradient staleness occurs when the local gradients are computed using an obsolete global model. 

To minimize the prediction error, i.e., $\underset{\boldsymbol{\upsilon}}{\text{\bf min }}\rho_j(\boldsymbol{\upsilon})$, we also apply the Adam optimizer and update the global model $\boldsymbol{\upsilon}^{(\phi+1)}$ as expressed in Eqs.~(\ref{eqn3d1})-(\ref{eqn3e}). This $\boldsymbol{\upsilon}^{(\phi+1)}$ is then pushed back to the CS-$j$, $\forall j \in \mathcal{J}$ for the next local learning process. The aforementioned process continues until the prediction error converges or a certain number of epoch time $T$ is reached. Then, we can predict $\mathbf{\hat Y}_{\emph{\mbox{j}}}, \forall j \in \mathcal{J}$ of training dataset $\mathbf{X}_{\emph{\mbox{j}}}, \forall j \in \mathcal{J}$ and new dataset $\mathbf{\hat X}_{\emph{\mbox{j}}}, \forall j \in \mathcal{J}$ at CS-$j$, $\forall j \in \mathcal{J}$ using $\boldsymbol{\upsilon}^*$ through Eq.~(\ref{eqn3f}). The algorithm for energy demand prediction using FEDL is shown in Algorithm~\ref{DDL-PC} in which the processes between Lines 9 and 11 is executed in the CSP.

\begin{algorithm}[]
	\caption{FEDL-Based Prediction Algorithm} \label{DDL-PC}
	
	\begin{algorithmic}[1] 
		
		\STATE Initialize $a_{j}$ and $\boldsymbol{\upsilon}^{(\phi)}$ when $\phi=0$
		
		\STATE Generate $\mathbf{X}_j, \forall j \in \mathcal{J}$
		
		\WHILE{$\phi \leq T \text{ {\bf and} } \rho_j(\boldsymbol{\upsilon}^{(\phi)}), \forall j \in \mathcal{J} \text{ do not converge}$}
		
		\FOR{$\forall j \in \mathcal{J}$}
		
		\STATE Learn $\mathbf{X}_j^1$ to obtain $\mathbf{Y}_j^L$ at layer-$L$ using $\boldsymbol{\upsilon}^{(\phi)}$
		
		\STATE Calculate $\rho_j(\boldsymbol{\upsilon}^{(\phi)})$ and $\nabla \boldsymbol{\upsilon}_j^{(\phi)}$
		
		\STATE Send $\nabla \boldsymbol{\upsilon}_j^{(\phi)}$ to the CSP for global model update
		
		\ENDFOR
		
		\STATE Compute current $\boldsymbol{\upsilon}^{(\phi)}$
		
		\STATE $\phi \leftarrow \phi+1$
		
		\STATE Update and send $\boldsymbol{\upsilon}^{(\phi)}$ back to $J$ CSs 
		
		\ENDWHILE
		
		\FOR{$\forall j \in \mathcal{J}$}
		
		\STATE Predict $\mathbf{\hat Y}_{\emph{\mbox{j}}}$ using $\mathbf{\hat X}_{\emph{\mbox{j}}}$ and $\boldsymbol{\upsilon}^*$ for the next-period energy requests
		
		\ENDFOR
		
	\end{algorithmic}
\end{algorithm}

\subsection{Clustering-Based Energy Demand Learning}
\label{sec:PF3}

Learning the dataset without considering the useful feature classification as the aforementioned approaches may produce biased energy demand prediction, especially when we combine imbalanced features and known labels in one dataset. Then, to obtain better prediction accuracy, we can group CSs into $K$ clusters before the learning process is performed. In this case, the clustering decision is determined by using location information of the CSs, i.e., the latitude and longitude. We utilize the constrained K-means algorithm~\cite{Bradley:2000} to perform the clustering process. In particular, we modify the constrained K-means algorithm to generate balance number of the CSs in each cluster through using minimum and maximum cluster size constraints. This consideration is to give a fairness of the learning process for the CSs in each cluster based on their deployment locations. Given $\mathcal{K} = \{1,\ldots,k,\ldots,K\}$ as the set of clusters and dataset $\mathbf{N}$ containing CS IDs and their locations $n_i, \forall i \in \mathcal{I}$, we aim to determine centroids (i.e., cluster centers) $C_k,\forall k \in \mathcal{K}$ such that the accumulation of squared distance between each $n_i$ and its closest centroid $C_k$ is minimized as optimization problem described below.
\begin{equation}
\label{eqn3j}
\begin{aligned}
\min_{\{\boldsymbol{\tau}, \mathbf{C}\}}\sum_{i \in \mathcal{I}}\sum_{k \in \mathcal{K}}\tau^k_i (n_i - C_k)^2,
\end{aligned}
\end{equation}
\begin{eqnarray}
\text{s.t.} \quad
\theta_k^{low} \leq \sum_{i \in \mathcal{I}}\tau^k_i \leq \theta_k^{high}, \forall k \in \mathcal{K}, \label{eqn3j1} \\
\sum_{k \in \mathcal{K}}\tau^k_i = 1, \forall i \in \mathcal{I}, \label{eqn3j2} \\
\tau^k_i \in \{0,1\}, \forall i \in \mathcal{I}, \forall k \in \mathcal{K}, \label{eqn3j3} \\
\theta^k_{low} \geq 0, \theta^k_{high} \geq 0,  \forall k \in \mathcal{K}, \label{eqn3j4}
\end{eqnarray}
where $\tau^k_i$ is a binary variable with $\tau^k_i=1$ representing that the location of CS-$i$ is the nearest to the centroid $C_k$ and $\tau^k_i=0$ otherwise. The constraints (\ref{eqn3j1}) guarantee that the CSs' locations in each cluster-$k$ are within predefined minimum $\theta^k_{low}$ and maximum $\theta^k_{high}$ thresholds. Additionally, the constraints (\ref{eqn3j2}) indicate that a location of CS-$i$ is clustered to the closest centroid only. To reach the optimal solution, we need to update the centroid $C_k^{(t)}$ at each iteration $t$. Specifically, in the cluster-$k$, we have
\begin{equation}
\label{eqn3m1}
\begin{aligned}
C_k^{(t+1)}=\frac{\underset{i \in \mathcal{I}}{\sum}\tau^{k,(t)}_{i}n_i}{\underset{i \in \mathcal{I}}{\sum}\tau^{k,(t)}_{i}},
\end{aligned}
\end{equation}
if $\theta^k_{low} \leq \underset{i \in \mathcal{I}}{\sum}\tau^k_i \leq \theta^k_{high}$ and 
\begin{equation}
\label{eqn3m1}
\begin{aligned}
C_k^{(t+1)}=C_k^{(t)},
\end{aligned}
\end{equation}
otherwise. The process terminates when $C_k^{(t+1)}=C_k^{(t)}, \forall k \in \mathcal{K}$, and thus we can obtain the optimal set $\mathcal{I}_k$ in the cluster-$k$. To this end, we can perform the centralized EDL or FEDL method to predict the energy demand for the CSs in each cluster independently. The algorithm for the clustering-based EDL using customized constrained K-means method is shown in Algorithm~\ref{ConsKMeans}.

\begin{algorithm}[]
	\caption{Clustering-Based EDL with Customized Constrained K-Means Algorithm} \label{ConsKMeans}
	
	\begin{algorithmic}[1] 
		
		\STATE Generate $\mathbf{N}$ containing $n_i, \forall i \in \mathcal{I}$
		
		\STATE Set $K$ and initialize random $C_k,\forall k \in \mathcal{K}$ when $t=0$
		
		\WHILE{$C_k^{(t+1)} \neq C_k^{(t)}, \forall k \in \mathcal{K}$}
		
		\STATE Solve optimization problem in Eq.~(\ref{eqn3j}) with the constraints in Eqs.~(\ref{eqn3j1})-(\ref{eqn3j4})
		
		\STATE Update $t+1$
		
		\FOR{$\forall k \in \mathcal{K}$}
		
		\IF{$\theta^k_{low} \leq \underset{i \in \mathcal{I}}{\sum}\tau^k_i \leq \theta^k_{high}$}
		
		\STATE $C_k^{(t+1)}=\frac{\underset{i \in \mathcal{I}}{\sum}\tau^{k,(t)}_{i}n_i}{\underset{i \in \mathcal{I}}{\sum}\tau^{k,(t)}_{i}}$
		
		\ELSE
		
		\STATE $C_k^{(t+1)}=C_k^{(t)}$
		
		\ENDIF
		
		\ENDFOR
		
		\ENDWHILE
		
		\STATE Generate the optimal set $\mathcal{I}_k$ in the cluster-$k$, $\forall k \in \mathcal{K}$
		
		\STATE Perform EDL or FEDL method using Algorithm~1 or~2, respectively, in the cluster-$k$, $\forall k \in \mathcal{K}$
		
	\end{algorithmic}
\end{algorithm}

\section{Performance Evaluation} 
\label{sec:PE}

\subsection{Dataset Pre-Processing and Evaluation Method} 
\label{subsec:DP}

To evaluate the performance of the proposed learning methods, we use the real data obtained from charging stations in Dundee city, the United Kingdom between 2017 and 2018~\cite{Dundee:2018}. In particular, the dataset has 65,601 transactions which include CS ID from 58 CSs, transaction ID for each CS, EV charging date, EV charging time, and consumed energy (in kWh) for each transaction. We use the first four information as the learning features, and the consumed energy as the learning label. Then, we classify CS ID, charging date, and charging time as categorical features. Specifically, we convert the charging date and charging time information into 7-day (i.e., $1,2,\ldots,7$) and 24-hour (i.e., $0,1,\ldots,23$) categories, respectively. In addition, each CS has the latitude and longitude information which will be used for clustering.


Then, we adopt the RMSE to show the prediction accuracy, i.e., prediction error, because we deal with the prediction of energy demand which is categorized as a regression prediction model, i.e., when the mapping function yields the continuous prediction outputs. Given $S$ transactions, the RMSE can be computed as follows:
\begin{equation}
\label{eqn3k}
\begin{aligned}
\mbox{RMSE} = \sqrt{\frac{1}{S}\sum_{s=1}^S (\omega_s - {\hat \omega}_s)^2},
\end{aligned}
\end{equation}
where $\omega_s$ and ${\hat \omega}_s$ are the actual and predicted energy demand for transaction $s$. 

\subsection{Experimental Setup} 
\label{subsec:ES}

\begin{figure}[!t]
	\centering
	\includegraphics[scale=0.31]{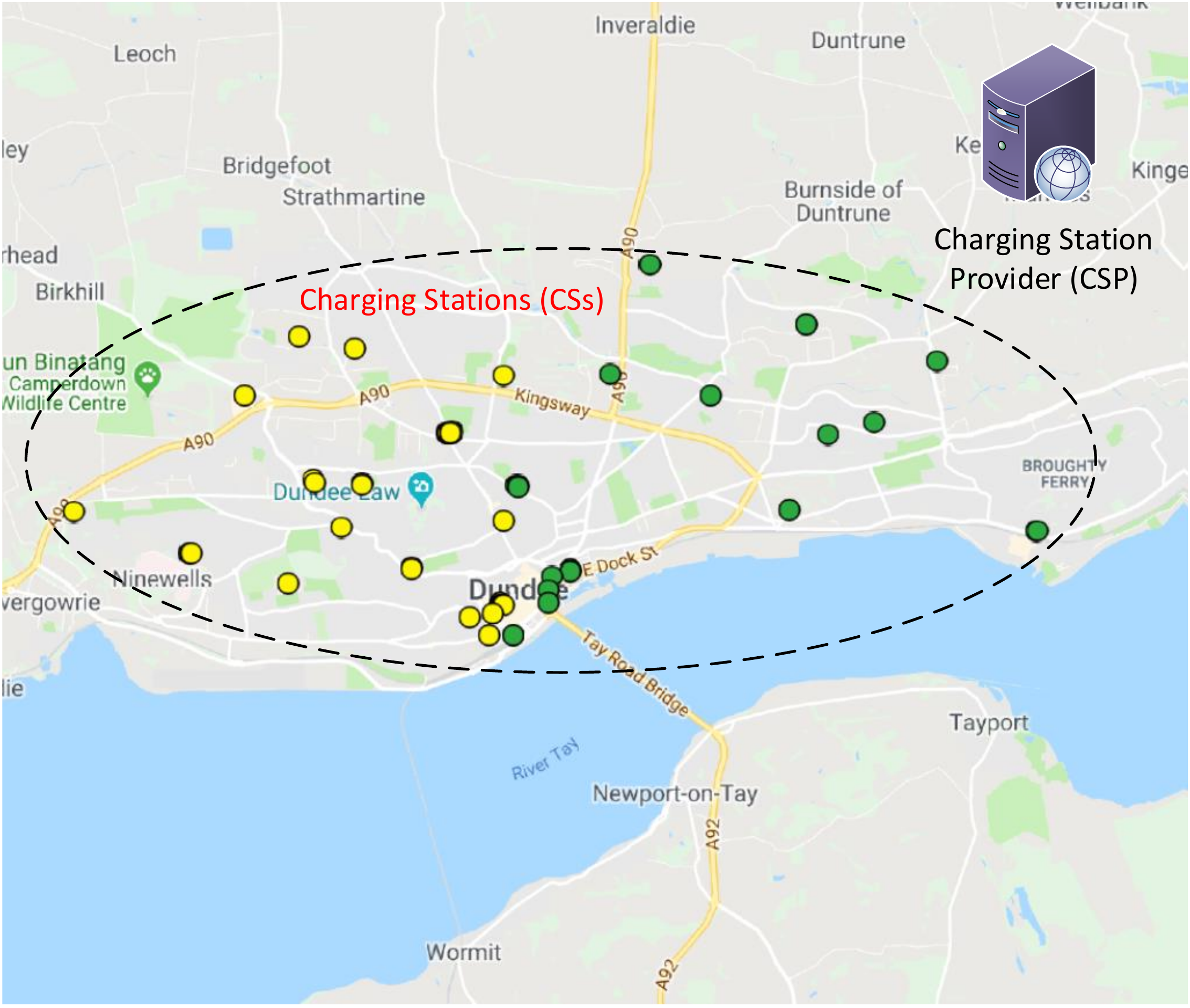}
	\caption{Charging station clustering.}
	\label{fig:Clustering_Topology}
\end{figure}

We evaluate the performance using \emph{TensorFlow CPU} in UTS shared cluster with configuration Intel Xeon E5-2687W v2 3.4GHz 8 cores 32GB RAM. We compare our proposed methods with some other conventional machine learning methods including decision tree (DT), random forest (RF), support vector regressor (SVR), k-neighbors regressor (KNR), stochastic gradient descent regressor (SGDR), and multi-layer perceptron regressor (MLPR)~\cite{Boutaba:2018}. We split the dataset into 80\%, 70\%, 60\%, as well as 50\% training dataset, and the rest of the portions for testing dataset. From the training dataset, we divide the number of transactions by $J$ training subsets, when FEDL is implemented. Each CS-$j$ runs the testing dataset for the upcoming energy demand prediction. For clustering, we define $K=2$, and thus we divide 58 CSs into 2 clusters as shown in Fig.~\ref{fig:Clustering_Topology}. For DNN, we use two hidden layers with 64 neurons per layer and one dropout layer with a fraction rate 0.15. We also apply the adaptive learning rate Adam optimizer with initial step size $0.01$ and \emph{tanh} function as the activation function.

\subsection{Simulation Results}
\label{subsec:SR}

Table~\ref{tab:contents2} shows the comparisons between baseline and proposed machine learning methods for various training set ratios. We first evaluate the prediction accuracy, i.e., RMSE, of the testing set when 80\% training set ratio is used. In particular, the RMSE obtained by the centralized EDL with clustering and the FEDL with clustering are 24.28\% and 24.63\% lower than those of the baseline methods, respectively. The reason is that the clustering approach can improve the prediction accuracy by grouping similar useful features and/or labels together in the same cluster~\cite{Li:2018}. As such, by clustering the CSs based on their locations, we can reduce the cost of biased prediction results over the whole dataset, which leads to the lower prediction error. For the FEDL and the centralized EDL without clustering, the RMSE performance is also always better than those of the baseline learning methods up to 23.51\%. Moreover, the gap between the FEDL as well as the centralized EDL without clustering and the proposed learning methods with clustering is only within 2\%. This is because the FEDL and the centralized EDL can deeply learn the useful features from the dataset. Specifically, the use of hyperparameter settings in the DNN, e.g., the number of hidden layers and neurons, the regularization methods, the activation functions, and the mini-batch size, contributes to the prediction accuracy improvement. Additionally, for the FEDL, it can learn the subset of the whole dataset independently at different workers, and achieve the average prediction with less variance and lower error in respect of the number of workers. 

\begin{table}[h!]
	\centering
	\caption{Testing RMSE of energy demand prediction for various learning methods and training set ratios}
	\begin{center}
		\begin{tabular}{ |c|c|c|c|c| } 
			\hline
			\multirow{2}{*}{\bf Learning approach} & \multicolumn{4}{|c|}{\bf Training set ratio} \\
			\cline{2-5}
			& {\bf 80\%} & {\bf 70\%} & {\bf 60\%} & {\bf 50\%} \\
			\hline
			\hline
			KNR & 7.18 & 7.71 & 7.57 & 7.67 \\
			\hline 
			MLPR & 6.57 & 6.62 & 6.90 & 6.53 \\
			\hline 
			SGDR & 6.55 & 6.57 & 6.54 & 6.54 \\
			\hline
			DT & 6.47 & 6.49 & 6.47 & 6.47 \\ 
			\hline
			SVR & 6.46 & 6.50 & 6.50 & 6.53 \\ 
			\hline
			RF & 6.35 & 6.66 & 6.88 & 6.80 \\ 
			\hline
			{\bf EDL} & 5.86 & 5.86 & 5.86 & 5.87 \\ 
			\hline
			{\bf FEDL} & 5.81 & 5.82 & 5.81 & 5.84 \\ 
			\hline
			{\bf EDL + Clustering} & {\bf 5.77} & {\bf 5.79} & 5.85 & 5.87 \\ 
			\hline
			{\bf FEDL + Clustering} & {\bf 5.76} & {\bf 5.78} & {\bf 5.78} & {\bf 5.83}  \\ 
			\hline
		\end{tabular}
	\end{center}
	\label{tab:contents2}
\end{table}

In constrast to the proposed learning methods, the baseline learning methods cannot learn the useful features and their correlations deeply. The reason is that they do not explore nonlinear transformations of the complex hidden features and multiple levels of processing layers as provided by the EDL and FEDL methods. For other training set ratios (i.e., 70\%, 60\%, and 50\%), we observe that the proposed learning methods still outperform all the baseline learning methods. In particular, the centralized EDL with clustering and the FEDL with clustering still have the lowest RMSE for training set ratio 70\%. Furthermore, the FEDL with clustering has the lowest RMSE for the rest of training set ratios. This interesting trend can provide useful information for the CSP to choose the suitable proposed learning methods in terms of stability, robustness, and flexibility.

\begin{figure}[!t]
	\centering
	\includegraphics[scale=0.3]{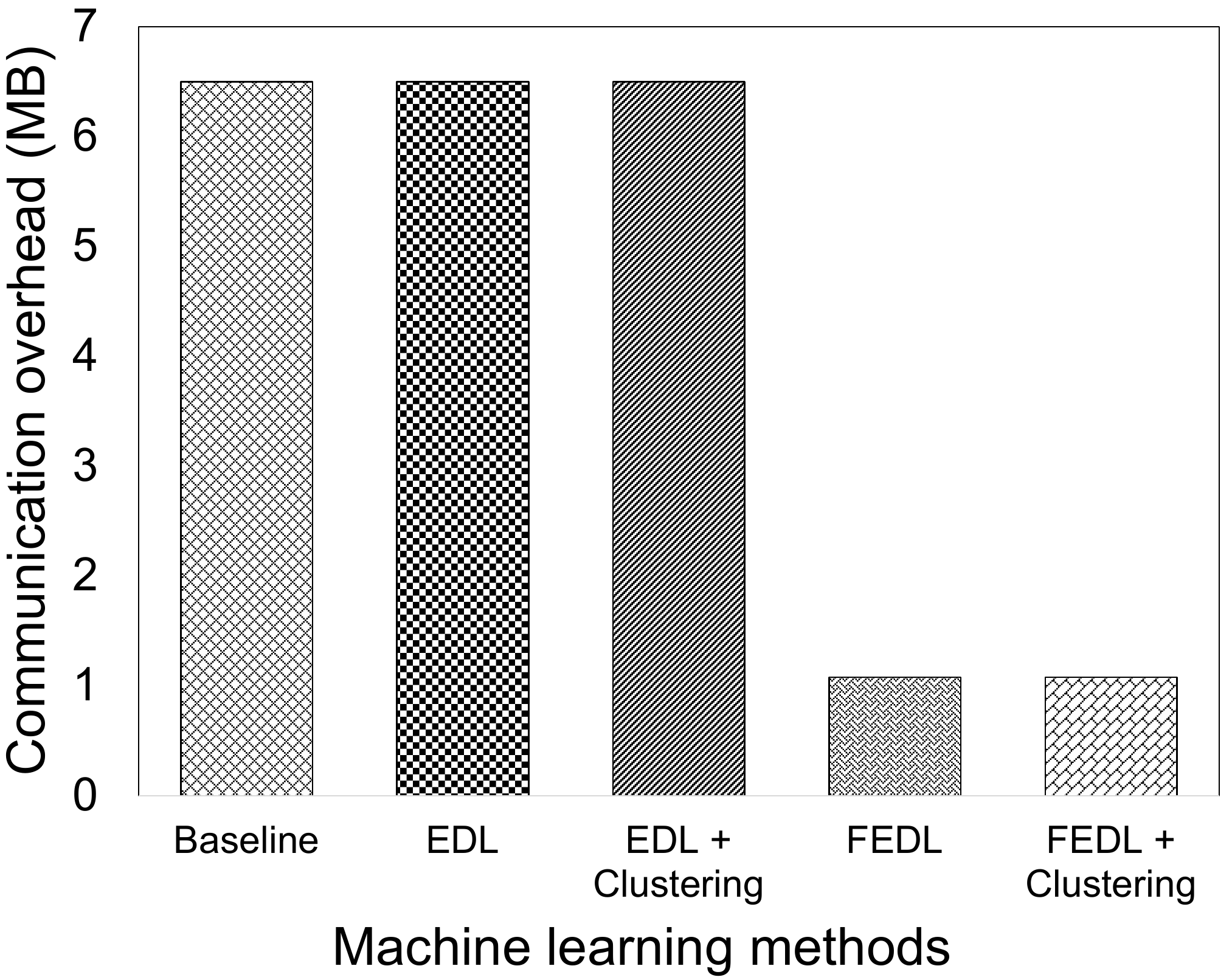}
	\caption{Communication overhead between baseline and proposed machine learning methods.}
	\label{fig:comm_over}
\end{figure}


In Fig.~\ref{fig:comm_over}, we show that the use of FEDL method can significantly reduce the communication overhead up to 83.4\% compared with the centralized methods including all baseline methods and the centralized EDL. This is because the CSP only requires to collect the trained models from the workers without sending any real dataset. This advantage aligns with the reduction of privacy issue for participating EVs and CSs.


\section{Conclusion}
\label{sec:Conc}
In this paper, we have proposed the novel machine learning-based approaches leveraging deep learning techniques to improve the energy demand prediction accuracy and reduce the communication overhead in the EV network. In the first approach, the CSP gathers information from all CS and then predict energy demand for the whole network using the deep learning technique. After that, we have introduced the federated energy demand learning-based method in which the energy demand learning can be performed at the CSs without disclosing the privacy of EVs and CSs. Furthermore, we have applied the clustering-based energy demand learning method for the CSs to further improve the energy demand prediction accuracy. Through the simulation results, we have demonstrated that our proposed approaches outperform other machine learning algorithms in terms of the prediction accuracy and communication overhead.



\ifCLASSOPTIONcaptionsoff
  \newpage
\fi

\vspace{0.7cm}

\clearpage
%
%
%

%
%
%


\begin{thebibliography}{100}
\Urlmuskip=0mu plus 1mu\relax
\bibliographystyle{IEEEtran}


\bibitem{Cnbc:2019}
The Growth of Electric Vehicles. Available Online: \url{https://www.cnbc.com/2018/05/30/electric-vehicles-will-grow-from-3-million-to-125-million-by-2030-iea.html}. Last Accessed on April 2019.


\bibitem{Tang:2014}
W.~Tang, S.~Bi, and Y.~J.~A.~Zhang, ``Online coordinated charging decision algorithm for electric vehicles without future information," \emph{IEEE Transactions on Smart Grid}, vol. 5, no. 6, pp. 2810-2824, Nov. 2014.

\bibitem{Lopes:2011}
J.~A.~P.~Lopes, F.~J.~Soares, and P.~M.~R.~Almeida, ``Integration of
electric vehicles in the electric power system," \emph{Proc. IEEE}, vol. 99,
no. 1, pp. 168–183, Jan. 2011.


\bibitem{You:2014}
P.~You and Z.~Yang, ``Efficient optimal scheduling of charging station
with multiple electric vehicles via V2V," in \emph{IEEE SmartGridComm}, Nov. 2014, pp. 716–721.

%



\bibitem{Majidpour:2015}
M.~Majidpour, \emph{et al.}, ``Fast prediction for sparse time series: demand forecast of EV charging stations for cell phone applications," \emph{IEEE Transactions on Industrial Informatics}, vol. 11, no. 1, pp. 242-250, Feb. 2015.


\bibitem{Chis:2017}
A.~Chis, J.~Lunden, and V.~Koivunen, ``Reinforcement learning-based plug-in electric vehicle charging with forecasted price," \emph{IEEE Transactions on Vehicular Technology}, vol. 66, no. 5, pp. 3674-3684, May 2017.

\bibitem{Fukushima:2018}
A.~Fukushima, \emph{et al.}, ``Prediction of energy consumption for new electric vehicle models by machine learning," \emph{IET Intelligent Transport Systems}, vol. 12, no. 9, pp. 1174-1180, Oct. 2018.

\bibitem{Ma:2017}
W.~Ma, V.~Gupta, and U.~Topcu, ``Distributed charging control of electric vehicles using online learning," \emph{IEEE Transactions on Automatic Control}, vol. 62, no. 10, pp. 5289-5295, Oct. 2017.

\bibitem{Lopez:2019}
K.~L.~Lopez, C.~Gagne, and M.~Gardner, ``Demand-side management using deep learning for smart charging of electric vehicles," \emph{IEEE Transactions on Smart Grid}, 2019.

\bibitem{He:2019}
Y.~He, K.~M.~Kockelman, and K.~A.~Perrine, ``Optimal locations of US fast charging stations for long-distance trip completion by battery electric vehicles," \emph{Journal of Cleaner Production}, vol. 214, pp.452-461, Mar. 2019.

\bibitem{Li:2018}
W.~Li, \emph{et al.}, "Implemented IoT-based self-learning home management system (SHMS) for Singapore," \emph{IEEE Internet of Things Journal}, vol. 5, no. 3, pp. 2212-2219, Jun. 2018.

\bibitem{Dundee:2018}
Electric Vehicle Charging Sessions Dundee. Available Online: \url{https://data.dundeecity.gov.uk/dataset/ev-charging-data}. Last Accessed on April 2019.

\bibitem{Roman:2011}
S.~Roman, \emph{et al.}, ``Regulatory framework and business models for charging plug-in electric vehicles: infrastructure, agents, and commercial relationships," \emph{Energy Policy}, vol. 39, no. 10, pp. 6360-6375, Oct. 2011.

\bibitem{Zhang2:2018}
C.~Zhang, P.~Patras, and H.~Haddadi, ``Deep learning in mobile and wireless networking: a survey," arXiv:1803.04311 [cs.NI], Sept. 2018.

%
\bibitem{Kingma:2015}
D.~Kingma and J.~Ba, ``Adam: a method for stochastic optimization," in \emph{ICLR 2015}, May 2015, pp. 1-15.

\bibitem{Dean:2012}
J.~Dean \emph{et al.}, ``Large scale distributed deep networks," in \emph{ACM NIPS 2012}, Dec. 2012, pp. 1223-1231.

\bibitem{Bradley:2000}
P.~S.~Bradley, K.~P.~Bennett, and A.~Demiriz, ``Constrained K-means clustering," \emph{Microsoft Research MSR-TR-2000-65}, May 2000.


%
%

\bibitem{Boutaba:2018}
R.~Boutaba, \emph{et al.}, ``A comprehensive survey on machine learning for networking: evolution, applications and research opportunities," \emph{Journal of Internet Services and Applications}, vol. 9, no. 1, pp. 1-99, Jun. 2018.

%
%
%
%
%

%
%
%
%
%
%
%
%
%
%
%
%
%
%
%
%
%
%
%
%
%
%
%
%
%
%
%
%
%
%
%
%
%
%
%
%
%
%
%
%
%
%
%
%
%
%
%
%
%
%
%
%
%
%
%
%
%
%
%
%
\end{thebibliography}
\end{document}